\documentclass[aps,prl,showpacs,twocolumn,amsmath,10pt,nofootinbib,superscriptaddress,floatfix]{revtex4}
\usepackage{graphicx,psfrag}
\usepackage{times}
\usepackage{color}
\usepackage{amsfonts,amssymb,stmaryrd,latexsym,amsmath}

\begin{document}
\title{Comment on ``Possibility of Deeply Bound Hadronic Molecules from Single Pion Exchange''}
\author{A.~A.~Filin}
\affiliation{Helmholtz-Institut f\"ur Strahlen- und
           Kernphysik and Bethe Center for Theoretical Physics, Universit\"at
          Bonn,  D--53115 Bonn, Germany}
\affiliation{Institute for Theoretical and Experimental Physics, B.\ Cheremushinskaya 25, 
          117218 Moscow, Russia}

\author{A. Romanov}
\affiliation{Institute for Theoretical and Experimental Physics, B.\ Cheremushinskaya 25, 
          117218 Moscow, Russia}

\author{V.\ Baru}
\affiliation{Institute for Theoretical and Experimental Physics, B.\ Cheremushinskaya 25, 
          117218 Moscow, Russia}
\affiliation{Institut f\"{u}r Kernphysik and J\"ulich Center
          for Hadron Physics, Forschungszentrum J\"{u}lich,
          D--52425 J\"{u}lich, Germany}

\author{C.\ Hanhart} 
\affiliation{Institut f\"{u}r Kernphysik and J\"ulich Center
          for Hadron Physics, Forschungszentrum J\"{u}lich,
          D--52425 J\"{u}lich, Germany}
\affiliation{Institute for Advanced Simulation,
          Forschungszentrum J\"{u}lich, D--52425 J\"{u}lich, Germany}

\author{Yu.~S.~Kalashnikova}
\affiliation{Institute for Theoretical and Experimental Physics, B.\ Cheremushinskaya 25, 
          117218 Moscow, Russia}

\author{A.~E.~Kudryavtsev}
\affiliation{Institute for Theoretical and Experimental Physics, B.\ Cheremushinskaya 25, 117218 Moscow, Russia}

\author{U.-G.~Mei\ss ner} 
\affiliation{Helmholtz-Institut f\"ur Strahlen- und
           Kernphysik and Bethe Center for Theoretical Physics, Universit\"at
          Bonn,  D--53115 Bonn, Germany}
\affiliation{Institut f\"{u}r Kernphysik and J\"ulich Center
          for Hadron Physics, Forschungszentrum J\"{u}lich,
          D--52425 J\"{u}lich, Germany}
\affiliation{Institute for Advanced Simulation,
          Forschungszentrum J\"{u}lich, D--52425 J\"{u}lich, Germany}

\author{A.~V.~Nefediev}
\affiliation{Institute for Theoretical and Experimental Physics, B.\ Cheremushinskaya 25, 
          117218 Moscow, Russia}


\maketitle



\newcommand{\boldpi}{\mbox{\boldmath $\pi$}}
\newcommand{\boldtau}{\mbox{\boldmath $\tau$}}
\newcommand{\boldT}{\mbox{\boldmath $T$}}
\newcommand{\gaprox}{$ {\raisebox{-.6ex}{{$\stackrel{\textstyle >}{\sim}$}}} $}
\newcommand{\saprox}{$ {\raisebox{-.6ex}{{$\stackrel{\textstyle <}{\sim}$}}} $}

In Ref.~\cite{close1} deeply bound systems of a pair of two open charm mesons,
for simplicity here called $D_\alpha$ and $D_\beta$, were predicted. It was
assumed that one of the two constituents, here $D_\beta$, had to have a large
width, dominated by the $S$-wave decay $D_\beta\to D_\alpha\pi$. Since this
large width implies a large coupling to pions in the transition potential, the
authors concluded, backed by a variational calculation~\cite{close1} and a
solution of a Schr\"odinger equation~\cite{close2}, that in some channels the
attraction in the $\bar D_\alpha D_\beta$ system must be sufficiently strong
to produce {\sl deeply} bound states. 
In this comment we present arguments that those states should not exist. 
In short our reasoning goes as follows: if $D_\beta$ has a significant 
$D_\alpha\pi$ decay width, this not only means that the $\bar D_\alpha
D_\beta$ interaction via pion exchange should be strong, but also that
in the dynamical equations for $\bar D_\alpha D_\beta$ scattering, the
$D_\beta$ width has to be included as well as the three--body cuts
due to $\bar D_\alpha D_\alpha \pi$ intermediate states.
That the former effect alone will already strongly distort the resonance shape
was discussed recently in Ref.~\cite{finitewidth}. For the case
at hand here we found from
an explicit calculation, which  reproduces the results of Ref.~\cite{close2}
once the approximations of that paper are imposed, that
 taking into account both aforementioned effects completely removes any signal
of bound states. Thus, we find that as soon as the $D_\beta\to D_\alpha \pi$
coupling is sufficiently strong to produce a bound state it is
at the same time {\it necessarily}
sufficiently strong to provide the state with such a large width that it
becomes unobservable. This connection is unavoidable for the interplay of
the various components is a consequence of 
unitarity~\cite{AAY}.

\begin{figure}[ht!]
\begin{center}
\psfrag{yyy}{\hspace*{-0.2cm}$d\sigma/dM_{\bar D^*D^*\pi}$ [arb. units]}
\psfrag{xxx}{\hspace*{-0.7cm}$M_{\bar D^*D^*\pi}-2M_{D^*}-m_\pi$ [MeV]}
\includegraphics[width=0.7\linewidth]{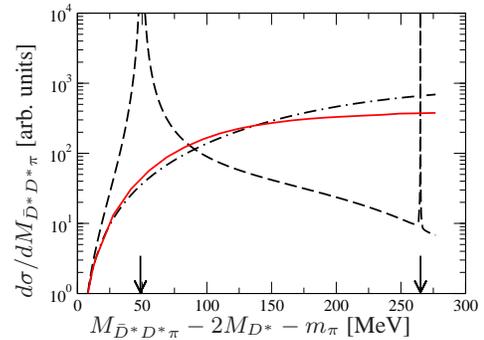}
\caption{Calculated $\bar D^*D^*\pi$ production cross sections with arbitrary
relative normalization.
Dashed line: using all approximations of Refs.~\cite{close1,close2}.
Dot-dashed line: adding simplified imaginary parts. Solid line: exact result.
For details see text.}
\label{fig}
\end{center}
\end{figure}

In order to make the arguments given more quantitative we now focus on the
example of a possible bound $\bar D_1D^*$ system in the isoscalar--vector channel.  To visualize our findings
we present the predicted $ \bar D^*D^* \pi$ invariant mass distribution assuming
all three particles to emerge from a point source.
To do the calculation we convolute the resulting transition matrix element
with the proper three--body phase space as well as the $D_1$ spectral
function. For the latter quantity in the phase space integration for
all calculations we use the correct expression, regardless what approximations
are used in the scattering equation.
 The parameters
underlying the calculation are 2427~MeV and 453~MeV for the mass and
the width of the $D_1$, respectively, where the latter is consistent 
with the strength of the potential given in Ref. \cite{close2}.
No form factors were used. In Fig.~\ref{fig} we show
as the dashed line the result of our calculation once all approximations of
Refs.~\cite{close1,close2} are imposed. The spectral distribution clearly 
shows the lowest two states as very sharp peaks. The binding energies are
227 MeV and 12 MeV, respectively, in  agreement with the claims of Ref.~\cite{close2}.
Then we add in the imaginary part of the potential as derived, but dropped in
Ref.~\cite{close1}, as well as a constant width for the $D_1$. This leads to
the dot--dashed result in Fig.~\ref{fig}. As one can see, both resonance
signals are completely gone. 
In addition, this simplified calculation is
already very close to the full result, as given by solid line,
 which is obtained by solving  the Lippmann-Schwinger type equation for $\bar D_1D^*$ system
 including the full $\bar D^*D^*\pi$ dynamics
(in particular, the $\bar D^*D^*\pi$ cut) with relativistic pions  as
well as the full energy dependence of the potential and of the $D_1$ width.
Moreover, going
to the full calculation does not introduce any new parameter, since all
individual contributions are linked through two-- and three--body
unitarity. Thus, the deeply bound hadronic molecules, advocated in
Refs.~\cite{close1,close2} do not exist.
\begin{acknowledgments}
Supported  by the  Helmholtz
Association,
DFG, 
EU HadronPhysics2, RFFI,
grants
NSh-4961.2008.2 \& NSh-4568.2008.2, 
``Rosatom'', ``Dynasty'',
 ICFPM, and FCT.
\end{acknowledgments}

%

\end{document}